\documentstyle[12pt]{article}
\topmargin - 1.0 true cm
\newread\epsffilein 
\newif\ifepsffileok 
\newif\ifepsfbbfound 
\newif\ifepsfverbose 
\newdimen\epsfxsize 
\newdimen\epsfysize 
\newdimen\epsftsize 
\newdimen\epsfrsize 
\newdimen\epsftmp 
\newdimen\pspoints 
\def\epsfbox#1{\global\def\epsfllx{72}\global\def\epsflly{72}%
 \global\def\epsfurx{540}\global\def\epsfury{720}%
 \def\lbracket{[}\def\testit{#1}\ifx\testit\lbracket
 \let\next=\epsfgetlitbb\else\let\next=\epsfnormal\fi\next{#1}}%
\def\epsfgetlitbb#1#2 #3 #4 #5]#6{\epsfgrab #2 #3 #4 #5 .\\%
 \epsfsetgraph{#6}}%
\def\epsfnormal#1{\epsfgetbb{#1}\epsfsetgraph{#1}}%
\def\epsfgetbb#1{%
\openin\epsffilein=#1
\ifeof\epsffilein\errmessage{I couldn't open #1, will ignore it}\else
 {\epsffileoktrue \chardef\other=12
 \def\do##1{\catcode`##1=\other}\dospecials \catcode`\ =10
 \loop
 \read\epsffilein to \epsffileline
 \ifeof\epsffilein\epsffileokfalse\else
 \expandafter\epsfaux\epsffileline:. \\%
 \fi
 \ifepsffileok\repeat
 \ifepsfbbfound\else
 \ifepsfverbose\message{No bounding box comment in #1; using defaults}\fi\fi
 }\closein\epsffilein\fi}%
\def\epsfclipstring{}
\def\epsfsetgraph#1{%
 \epsfrsize=\epsfury\pspoints
 \advance\epsfrsize by-\epsflly\pspoints
 \epsftsize=\epsfurx\pspoints
 \advance\epsftsize by-\epsfllx\pspoints
 \epsfxsize\epsfsize\epsftsize\epsfrsize
 \ifnum\epsfxsize=0 \ifnum\epsfysize=0
 \epsfxsize=\epsftsize \epsfysize=\epsfrsize
 \epsfrsize=0pt
 \else\epsftmp=\epsftsize \divide\epsftmp\epsfrsize
 \epsfxsize=\epsfysize \multiply\epsfxsize\epsftmp
 \multiply\epsftmp\epsfrsize \advance\epsftsize-\epsftmp
 \epsftmp=\epsfysize
 \loop \advance\epsftsize\epsftsize \divide\epsftmp 2
 \ifnum\epsftmp>0
 \ifnum\epsftsize<\epsfrsize\else
 \advance\epsftsize-\epsfrsize \advance\epsfxsize\epsftmp \fi
 \repeat
 \epsfrsize=0pt
 \fi
 \else \ifnum\epsfysize=0
 \epsftmp=\epsfrsize \divide\epsftmp\epsftsize
 \epsfysize=\epsfxsize \multiply\epsfysize\epsftmp
 \multiply\epsftmp\epsftsize \advance\epsfrsize-\epsftmp
 \epsftmp=\epsfxsize
 \loop \advance\epsfrsize\epsfrsize \divide\epsftmp 2
 \ifnum\epsftmp>0
 \ifnum\epsfrsize<\epsftsize\else
 \advance\epsfrsize-\epsftsize \advance\epsfysize\epsftmp \fi
 \repeat
 \epsfrsize=0pt
 \else
 \epsfrsize=\epsfysize
 \fi
 \fi
 \ifepsfverbose\message{#1: width=\the\epsfxsize, height=\the\epsfysize}\fi
 \epsftmp=10\epsfxsize \divide\epsftmp\pspoints
 \vbox to\epsfysize{\vfil\hbox to\epsfxsize{%
 \ifnum\epsfrsize=0\relax
 \includegraphics{#1}%
 \else
 \epsfrsize=10\epsfysize \divide\epsfrsize\pspoints
 \includegraphics{#1}%
 \fi
 \hfil}}%
\global\epsfxsize=0pt\global\epsfysize=0pt}%
 {\catcode`\%=12 \global\let\epsfpercent=
\long\def\epsfaux#1#2:#3\\{\ifx#1\epsfpercent
 \def\testit{#2}\ifx\testit\epsfbblit
 \epsfgrab #3 . . . \\%
 \epsffileokfalse
 \global\epsfbbfoundtrue
 \fi\else\ifx#1\par\else\epsffileokfalse\fi\fi}%
\def\epsfempty{}%
\def\epsfgrab #1 #2 #3 #4 #5\\{%
\global\def\epsfllx{#1}\ifx\epsfllx\epsfempty
 \epsfgrab #2 #3 #4 #5 .\\\else
 \global\def\epsflly{#2}%
 \global\def\epsfurx{#3}\global\def\epsfury{#4}\fi}%
\def\epsfsize#1#2{\epsfxsize}
\newcommand{\AmS}{{\protect\the\textfont2
 A\kern-.1667em\lower.5ex\hbox{M}\kern-.125emS}}

\newcommand{\beq}{\begin{equation}}
\newcommand{\eeq}{\end{equation}}

\def\sss{\scriptscriptstyle}
\newcommand{\bd}{B_d^0}
\newcommand{\bdbar}{\overline{B_d^0}}
\newcommand{\bs}{B_s^0}
\newcommand{\bsbar}{\overline{B_s^0}}

\def\barp{{\raise.35ex\hbox{${\sss (}$}}---{\raise.35ex\hbox{${\sss )}$}}}
\def\bdbarp{\hbox{$B_d$\kern-1.4em\raise1.4ex\hbox{\barp}}}
\def\bsbarp{\hbox{$B_s$\kern-1.4em\raise1.4ex\hbox{\barp}}}
\def\dbarp{\hbox{$D$\kern-1.1em\raise1.4ex\hbox{\barp}}}

\def\dbar{{\overline{D^0}}}
\def\xD{x_{\sss D}}
\def\ddbar{$D^0$-$\dbar$}
\def\bbar{{\overline{B^0}}}
\def\bbbar{$B^0$-$\bbar$}
\def\bdbdbar{$\bd$-$\bdbar$}
\def\bsbsbar{$\bs$-$\bsbar$}
\def\rly#1{\mathrel{\raise.3ex\hbox{$#1$\kern-.75em\lower1ex\hbox{$\sim$}}}}
\def\lsim{\rly<}
\def\ks{K_{\sss S}}

\def\etal{{\it et al.\/}}

\def\npb#1#2#3{{\it Nucl.\ Phys.} {\bf B#1} (19#2) #3}
\def\plb#1#2#3{{\it Phys.\ Lett.} {\bf #1B} (19#2) #3}

\def\prd#1#2#3{{\it Phys.\ Rev.} {\bf D#1} (19#2) #3}
\def\pr#1#2#3{{\it Phys.\ Rev.} {\bf #1} (19#2) #3}

\def\prl#1#2#3{{\it Phys.\ Rev.\ Lett.} {\bf #1} (19#2) #3}

\def\zpc#1#2#3{{\it Zeit.\ Phys.} {\bf C#1} (19#2) #3}

\begin{document}
\begin{flushright}
UdeM-GPP-TH-96-39 \\
November 1996
\end{flushright}
\vskip.5truecm
\begin{center}
{\Large \bf
\centerline
{CP Violation and Heavy Hadrons}}
\vspace*{1.0cm}
 {\large David London}$\footnote{Presented at the 2nd International
Conference on Hyperons, Charm and Beauty Hadrons, Montr\'eal,
Canada, August 1996.}$
\vskip0.2cm
  Laboratoire de physique nucl\'eaire, Universit\'e de
Montr\'eal \\
    C.P. 6128, succ. centre-ville, Montr\'eal, QC, Canada
H3C 3J7\\
\vskip1.0cm
{\Large Abstract\\}
\parbox[t]{\textwidth}{
\indent
I review several topics involving CP violation with heavy hadrons. In
particular, I discuss (i) Hyperons: CP violation in the decay $\Lambda^0
\to p \pi^-$, (ii) Charm: indirect CP violation in the $D^0$ system, both
within and beyond the SM, and (iii) Beauty: indirect CP violation in the
neutral $B$-meson system beyond the SM.
}
\end{center}


\section{Introduction}

CP violation is one of the most intriguing mysteries in particle physics.
To date, it has been unambiguously observed only in
$K^0$-${\overline{K^0}}$ mixing. According to the standard model (SM), CP
violation is due to a complex phase in the Cabibbo-Kobayashi-Maskawa (CKM)
matrix. However, since this one parameter is included to ``explain'' only
one experimental measurement, we can hardly claim to understand the origin
of CP violation. If we want to go further, we will need to see CP violation
outside of the kaon system.

Still, even if CP violation is observed elsewhere, we will want to know the
answers to a number of questions:

\begin{itemize}

\item Can this new CP violation be explained by the phase of the CKM
matrix? In other words, we need to know the SM predictions for CP violation
outside the kaon system.

\item If not, what new physics could be responsible? I.e., we need to know
the beyond-the-SM predictions for CP violation.

\item Can we identify the new physics? That is, can we distinguish among
the various new physics possibilities?

\end{itemize}

In this talk, I will discuss several possibilities for the observation of
CP violation outside the kaon system. However, I should stress that the
subject of CP violation with heavy hadrons is vast. Thus, in light of time
constraints, I will restrict my discussion to 3 topics, which have been
inspired by the title of this conference:

\begin{enumerate}

\item Hyperons: CP violation in $\Lambda^0 \to p \pi^-$,

\item Charm: Indirect CP violation with neutral $D$ mesons, both within and
beyond the SM,

\item Beauty: Indirect CP violation with neutral $B$ mesons beyond the SM.

\end{enumerate}

There are numerous subjects which I don't have the time to discuss. These
include: the electric dipole moment of the neutron, triple products, direct
CP violation with $D$ mesons (both within and beyond the SM), charmed
baryons, direct CP violation in the $B$ system (within and beyond the SM),
indirect CP violation in the neutral $B$ system within the SM, $B$ baryons,
etc.

\section{Hyperons}

Consider the decay $\Lambda^0 \to p \pi^-$. This is a complex system --
the final state can be in an $s$-wave (parity-violating amplitude) or
$p$-wave (parity-conserving), and can have isospin $I=1/2$ or $3/2$. The
most general Lorentz-invariant amplitude for this decay can be written
\cite{Marshak}
\beq
{\cal M} = G_{\sss F} \, m_\pi^2 \, {\bar u}_p (A - B \gamma_5) \,
u_\Lambda~. 
\eeq
We define $\hat\sigma$ to be the polarization of the $\Lambda$, $\vec q$ to
be the momentum of the proton, and
\begin{eqnarray}
s \! &\equiv& \! A ~, \\
p \! &\equiv& \! \left( {|{\vec q}| \over E_p + M_p} \right) B ~. 
\end{eqnarray}
The differential cross section for this process is a complicated function
of $s$, $p$, and the spins of the $\Lambda$ and the $p$. However, if the
proton polarization is not measured, it can be written 
\beq
{d \Gamma \over d \Omega} \sim 1 + \alpha \, {\hat q} \cdot {\hat \sigma}
~, 
\eeq
where
\beq
\alpha = 2 { {\rm Re} \, s^* p \over |s|^2 + |p|^2 }~.
\eeq

Now compare $\Lambda^0 \to p \pi^-$ with ${\overline{\Lambda^0}} \to {\bar
p} \pi^+$. If CP is a good symmetry, $\alpha = - {\bar \alpha}$. Therefore
we define the CP-violating asymmetry \cite{Donoghue}
\beq
A \equiv {\alpha + {\bar\alpha} \over \alpha - {\bar\alpha}}~.
\eeq
(If the proton polarization is measured, there are additional CP-violating
asymmetries \cite{Donoghue}.) We would like to calculate the prediction for
this quantity in the SM. To this end, we separate $s$ and $p$ into $I=1/2$
and $3/2$ pieces \cite{Overseth}:
\begin{eqnarray}
s \! &=& \! -\sqrt{{2\over 3}} \, s_1 \,  
e^{i(\delta_1^s + \phi_1^s)}
+ \sqrt{{1\over 3}} \, s_3 \, e^{i(\delta_3^s + \phi_3^s)} ~, \\
p \! &=& \! -\sqrt{{2\over 3}} \, p_1 \, 
e^{i(\delta_1^p + \phi_1^p)}
+ \sqrt{{1\over 3}} \, p_3 \, e^{i(\delta_3^p + \phi_3^p)} ~,
\end{eqnarray}
where the $\delta$'s are the strong phases and the $\phi$'s are the weak
(CKM) phases. The key point is that all these quantities, apart from the
weak phases, have been measured experimentally \cite{expt,PDG96}. Plugging
in the measured values, we obtain
\begin{eqnarray}
A(\Lambda^0_-) \!\! &=& \!\! 0.13 \sin(\phi_1^p - \phi_1^s) +
0.001 \sin (\phi_1^p - \phi_3^s) \nonumber \\
&~& -~0.0024 \sin (\phi_3^p - \phi_1^s) ~. 
\end{eqnarray}
To finish the job, we now need to calculate the weak phases for the various
spin and isospin amplitudes.

In order for there to be CP violation at all, there must be at least two
amplitudes with different CKM phases which contribute to the decay process.
If not, all the $\phi$'s are equal, and the asymmetry vanishes. For the
decay $\Lambda^0 \to p \pi^-$, there are, in fact, several such amplitudes.
The tree contributions have CKM phase $V_{ud}^*V_{us}$, while the penguin
contributions have a phase $V_{ud}^*V_{us}$, $V_{cd}^*V_{cs}$, or
$V_{td}^*V_{ts}$, depending on the internal quark in the loop. (Using the
unitarity of the CKM matrix, the $V_{cd}^*V_{cs}$ piece can be eliminated
in favour of $V_{ud}^*V_{us}$ and $V_{td}^*V_{ts}$.) At the quark level,
one uses the effective Hamiltonian \cite{Buras}
\beq
H_{\sss W}^{\sss SM} = {G_{\sss F} \over \sqrt{2}} \, 
V_{ud}^* V_{us} \sum_i c_i(\mu) Q_i(\mu) + h.c.,
\eeq
where the $c_i(\mu)$ are the Wilson coefficients and the $Q_i(\mu)$ are the
4-quark operators. Using the renormalization group, the $c_i$'s can be
calculated; in the Wolfenstein parametrization the CP-violating phase is
\beq
{\rm Im} \left( {V_{td}^*V_{ts} \over V_{ud}^*V_{us}} \right) = A^2
\lambda^4 \eta \le 0.001~.
\eeq

So far, so good. However, the quarks now have to be put into hadrons. That
is, we have to calculate the hadronic matrix elements
\beq
\langle p \pi | H_{\sss W}^{\sss SM} | \Lambda^0 \rangle^{\sss I}_l = 
{\rm Re} \, M_l^{\sss I} + {\rm Im} \, M_l^{\sss I}~,
\eeq
where
\beq
\phi_l^{\sss I} \approx {{\rm Im} \, M_l^{\sss I} \over {\rm Re} \,
M_l^{\sss I}}~.
\eeq
Unfortunately, we don't know how to calculate these hadronic matrix
elements, and it is here that a large uncertainty enters the SM prediction.
The best we can do is to use the vacuum saturation approximation, which we
know is unreliable, since it cannot reproduce the $\Delta I=1/2$ rule. In
this case we find \cite{vacsat,HePak}
\beq
A(\Lambda^0_-) \approx - ({\hbox{1 -- 5}}) \times 10^{-5}.
\eeq
(The E871 experiment \cite{E871} expects to reach a sensitivity of about
$10^{-4}$ on $A(\Lambda^0_-) + A(\Xi^-_-)$, where $\Xi^-_-$ corresponds to
the process $\Xi^- \to \Lambda^0 \pi^-$. The SM prediction for $A(\Xi^-_-)$
is $-({\hbox{1 -- 10}})\times 10^{-5}$ \cite{HePak}, so an asymmetry might
just barely be measurable.) The main point here is that, due to hadronic
uncertainties, the SM prediction for this asymmetry is very imprecise -- we
really only know its order of magnitude.

How does the prediction for this asymmetry change in the presence of
physics beyond the SM? New physics can affect the asymmetry only if there
are new decay amplitudes. One way to analyze this is to use an effective
lagrangian involving all possible 4-quark operators, and calculate their
contributions to $A(\Lambda^0_-)$ including constraints from $\epsilon$ and
$\epsilon'/\epsilon$ \cite{Valencia}. If the scale of new physics is less
than 8 TeV, one finds that certain operators can give an asymmetry
$A(\Lambda^0_-) \sim 10^{-4}$. On the other hand, in most {\it models} of
new physics the new operators are not all independent, so the effective
lagrangian analysis may not tell the whole story. For example, in the
Weinberg model, one finds $A(\Lambda^0_-) \sim -2.5 \times 10^{-5}$, and
the ``isoconjugate'' left-right symmetric model gives $A(\Lambda^0_-) \sim
-1.1 \times 10^{-5}$ \cite{Donoghue,HePak,Chang}. In other words, despite
the effective-lagrangian analysis, in specific models it seems difficult to
obtain larger asymmetries than in the SM.

To sum up: the SM predictions for CP violation in hyperon decays have large
uncertainties. In the presence of new physics, the CP-violating asymmetries
may be larger than in the SM, but the calculations are both uncertain and
model-dependent. It may be possible to observe such asymmetries
experimentally, but (i) even if they are observed, it may not be clear
whether or not new physics is involved, and (ii) even if new physics is
involved, it will be very difficult to identify it. All in all, this is a
very messy system.

\section{Charm}

In order to get indirect CP violation in neutral $D$-meson decays, one
needs a final state $f$ to which both $D^0$ and $\dbar$ can decay. Then the
interference of the two amplitudes $D^0 \to f$ and $D^0 \to \dbar \to f$
can lead to CP violation. Obviously, for this to occur, one needs \ddbar\
mixing.

In the SM, the short-distance contributions to \ddbar\ mixing are due to
box diagrams with internal $d$, $s$ and $b$ quarks. The calculation of
these yields \cite{Datta}
\beq
\xD \equiv {\Delta M_{\sss D} \over \Gamma_{\sss D}} \sim 10^{-6} ~.
\eeq
However, since all particles in the loops are light compared to the weak
scale, the short-distance calculation is unreliable -- long-distance
effects can be important. Two estimates of these long-distance
contributions have been done. Using intermediate dispersive contributions,
one finds $\xD \lsim 6 \times 10^{-5}$ \cite{HewTak}, and heavy quark
effective theory gives $\xD \sim 6 \times 10^{-6}$ \cite{DmixHQET}. The
upshot is that \ddbar\ mixing is tiny in the SM: the $D$ meson will almost
always decay before mixing. Thus, the SM predicts essentially no indirect
CP violation in the neutral $D$-meson system. 

On the other hand, the present experimental limit on \ddbar\ mixing, coming
from $D^0 \to K^+ \pi^-$, is well above the SM prediction \cite{Dmixexpt}:
\beq
\label{exptDmix}
x_{\sss D}^{expt} < 0.083~.
\eeq
So there is plenty of room for new physics to contribute to such mixing.
And in fact, there are many models of new physics which do just that:
\newpage
\begin{enumerate}

\item Fourth Generation \cite{4gen}: Box diagrams with internal $b'$ quarks
contribute to \ddbar\ mixing.

\item $Z$-mediated FCNC's \cite{D-ZFCNC}: If the $u$- and $c$-quarks mix
with a left-handed singlet up-type quark, flavour-changing neutral currents
(FCNC's) of the $Z$ are induced. These FCNC's lead to \ddbar\ mixing at
tree level.

\item Multi-Higgs-Doublet Models: If one imposes natural flavour
conservation (NFC), then there are new contributions to the mixing through
box diagrams with internal charged-Higgses and $b$ quarks \cite{MHDM-NFC}.
These are important for large values of $\tan\beta$. If NFC is not imposed,
there will be flavour-changing couplings of the neutral Higgses, leading to
tree-level contributions to \ddbar\ mixing \cite{MHDM-noNFC}.

\item Supersymmetry with Quark-Squark Alignment \cite{Nir}: In this class
of non-minimal SUSY models, box diagrams with internal gluinos and $u$- and
$c$-squarks contribute to \ddbar\ mixing.

\item Light Scalar Leptoquarks \cite{Leurer}: Here the contributions to the
mixing come from box diagrams with internal leptons and leptoquarks.

\end{enumerate}
In all cases, for certain choices of the new-physics parameters, these
models can yield values of $\xD$ up to the experimental limit. (In fact, if
the mixing is as large as the limit of Eq.~\ref{exptDmix}, the analysis
leading to this limit may be invalidated \cite{Blaylock}.) Also, in all
cases new phases are introduced into \ddbar\ mixing.

The key point here is that, if we see large \ddbar\ mixing, this is a clear
signal of new physics. If such a mixing is observed, it is possible that we
may also find indirect CP violation. In order to do so, we need a final
state $f$ to which both $D^0$ and $\dbar$ can decay. One possibility is to
look at doubly-Cabbibo-suppressed (DCS) $D^0$ decays, for example into the
final state $\pi^- K^+$. The decay $D^0 (t) \to \pi^- K^+$ interferes,
through mixing, with the Cabibbo-allowed (CA) decay $\dbar (t) \to \pi^-
K^+$. ($D^0(t)$ and $\dbar (t)$ are the time-evolved states which at $t=0$
were $D^0$ and $\dbar$, respectively.) CP violation will be indicated (with
a caveat, to be discussed below) by an asymmetry $a_{\sss CP}^{\sss DCS}$
in the rates for these two processes. This CP asymmetry measures the
relative phase of the two amplitudes:
\begin{eqnarray}
a_{\sss CP}^{\sss DCS} \!\! &=& \!\! {\rm Im}\left[ {(D^0 \to \pi^-
K^+) \over (D^0 \to \dbar) \, (\dbar \to \pi^- K^+)} \right] \nonumber \\
&=& \!\! {\rm Im} \left[ {V_{cd} V_{us}^* \, e^{i \delta_{\sss DCS}}
\over \phi_{\sss M} V_{cs}^* V_{ud} \, e^{i \delta_{\sss CA}} } \right] \\
&=& \!\! - \left| {V_{cd} V_{us} \over V_{cs} V_{ud}} \right| 
\sin ( \phi_{\sss M} + \delta_{\sss CA} - \delta_{\sss DCS} )~, \nonumber
\end{eqnarray}
where $\phi_{\sss M}$ is the (weak) phase of \ddbar\ mixing, the $\delta$'s
are strong phases, and I have used the Wolfenstein parametrization, in
which the CKM matrix elements involving the first two generations are
essentially real. There are two points which should be noted here. First,
the asymmetry is small [$O(\lambda^2)$]. This is because the two
interfering decay amplitudes are not of comparable size. Second, this
asymmetry depends on the unknown strong phases -- in fact, even if
$\phi_{\sss M} = 0$, the asymmetry would still be nonzero. (Obviously, this
would not be a signal of CP violation, but is an example of how the strong
phases can ``fake'' CP violation.) 

Fortunately, it is possible to disentangle the weak and strong phases by
also looking at the CP-conjugate asymmetry ${\bar a}_{\sss CP}^{\sss DCS}$,
i.e.\ comparing the rates for $\dbar (t) \to \pi^+ K^-$ and $D^0 (t) \to
\pi^+ K^-$:
\beq
{\bar a}_{\sss CP}^{\sss DCS} = - \left| {V_{cd} V_{us} \over V_{cs}
V_{ud}} \right| 
\sin ( -\phi_{\sss M} + \delta_{\sss CA} - \delta_{\sss DCS}).
\eeq
By using both asymmetries, one can determine $\phi_{\sss M}$ and
$\delta_{\sss CA} - \delta_{\sss DCS}$, up to discrete ambiguities.

On the other hand, one can avoid all dependence on strong phases, and get a
larger asymmetry, by choosing a singly-Cabibbo-suppressed (SCS) decay such
as $\dbarp \to \pi^+\pi^-$. In this case,
\begin{eqnarray}
a_{\sss CP}^{\sss SCS} \!\! &=& \!\! {\rm Im}\left[ {(D^0 \to
\pi^+\pi^-) \over (D^0 \to \dbar) \, (\dbar \to \pi^+\pi^-)} \right]
\nonumber \\ 
&=& \!\! {\rm Im} \left[ {V_{cd} V_{ud}^* \, e^{i \delta_{\sss SCS}}
\over \phi_{\sss M} V_{cd}^* V_{ud} \, e^{i \delta_{\sss SCS}} } \right] \\
&=& \!\! - \sin ( \phi_{\sss M} ) ~. \nonumber
\end{eqnarray}
Since the two decay amplitudes are the same size, the asymmetry can be
quite large, considerably larger, in fact, than was the case for 
doubly-Cabbibo-suppressed $D^0$ decays. Furthermore, since a CP-eigenstate
final state is used, there is no strong phase dependence in the asymmetry.

Finally, the number of $D$'s needed to measure a CP-asymmetry is
proportional to $1/(BR(\dbarp \to f) a_{\sss CP}^2$). Because of this, it
is easier to look for CP violation using singly-Cabibbo-suppressed $D$
decays -- although the branching ratio is smaller, the asymmetry is
considerably larger. On the other hand, doubly-Cabbibo-suppressed $D^0$
decays are more useful in the search for \ddbar\ mixing.

\section{Beauty}

The $B$ system is the most promising place to look for CP violation outside
of the kaon system. The SM predicts large CP-violating asymmetries in
certain decays of neutral $B$ mesons. I will not give more than a cursory
review of the SM predictions for indirect CP violation in the $B$ system,
as this subject is covered in more detail elsewhere in these proceedings
\cite{Oliver}.

The phase information of the CKM matrix can be displayed elegantly using
the so-called unitarity triangle (Fig.~1). The 3 internal angles, $\alpha$,
$\beta$ and $\gamma$, can be probed through indirect CP violation in the
$B$ system. As in the charm system, such CP violation occurs through the
interference of the two amplitudes $B^0 \to f$ and $B^0 \to \bbar \to f$,
where $f$ is a final state to which both $B^0$ and $\bbar$ can decay. The
angles $\alpha$, $\beta$ and $\gamma$ can be measured through CP violation
in the decays $\bdbarp \to \pi^+ \pi^-$, $\bdbarp \to \Psi \ks$ and
$\bsbarp \to D_s^\pm K^\mp$, respectively. These angles are already
somewhat constrained by present experimental data: within the SM, one has
$-0.90 \leq \sin 2\alpha \le 1.0$, $0.32 \leq \sin 2\beta \le 0.94$, and
$0.34 \leq \sin^2 \gamma \le 1.0$ \cite{AliLon}.

\begin{figure}
\vskip -1.0truein
\centerline{\epsfxsize 3.5 truein \epsfbox {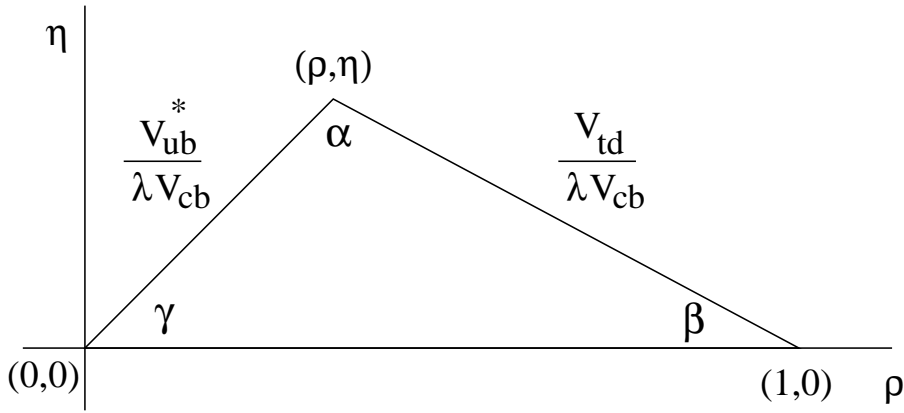}}
\vskip -1.1truein
\caption{The unitarity triangle. The angles $\alpha$, $\beta$ and $\gamma$
can be measured via CP violation in the $B$ system.}
\label{triangle}
\end{figure}

Through a measurement of these CP asymmetries, the presence of new physics
can be detected \cite{GroLon}. This can be done in 3 ways:
\begin{enumerate}

\item The relation $\alpha+\beta+\gamma=\pi$ is violated.

\item Although $\alpha+\beta+\gamma=\pi$, one finds values for the CP
phases which are outside of the SM predictions.

\item The CP angles measured are consistent with the SM predictions, and
add up to $180^\circ$, but are inconsistent with the measurements of the
{\it sides} of the unitarity triangle.

\end{enumerate}
In any of these cases, we will want to identify the type of new physics
which is responsible.

As in the $D$ system, new physics enters principally through \bbbar\
mixing, since there are no models of new physics which contribute
significantly to $B$ decays. The key question concerns the phase of the new
contributions. If the phase of the new-physics contribution is the same as
that of the SM, then the CP asymmetries will be unchanged from the SM
predictions. However, since the $V_{td}/\lambda V_{cb}$ side of the
unitarity triangle is extracted from the measured value of \bbbar\ mixing,
it {\it will} differ from its SM value. Thus, the new physics will be
detected via item (3) above -- the angles and the sides of the triangle
will be inconsistent with one another. On the other hand, if the phase of
the new-physics contribution is different from that in the SM, the CP
asymmetries will themselves be changed, and the new physics can be detected
via any of items (1)-(3).

There are a number of models of new physics which can contribute to \bbbar\
mixing \cite{GroLon}:
\begin{enumerate}

\item Fourth Generation: Box diagrams with internal $t'$ quarks contribute
to \bbbar\ mixing. There are new phases.

\item $Z$-mediated FCNC's: The mixing of the ordinary down-type quarks with
a left-handed singlet down-type quark induces flavour-changing couplings of
the $Z$. These FCNC's lead to \bbbar\ mixing at tree level, with new
phases.

\item Multi-Higgs-Doublet Models with NFC: There are new contributions to
the mixing through box diagrams with internal charged-Higgses and $t$
quarks. The phase is the same as in the SM.

\item Multi-Higgs-Doublet Models without NFC: If NFC is not imposed, there
will be flavour-changing couplings of the neutral Higgses, leading to
tree-level contributions to \bbbar\ mixing. There can be new phases.

\item Minimal Supersymmetry: There are numerous new contributions to
$B^0$-$\bbar$ mixing, through box diagrams with internal ordinary and
supersymmetric particles. In minimal SUSY, all contributions have the same
phase as the SM.

\item Non-minimal Supersymmetry: In non-minimal SUSY models, the new box
diagrams can have different phases than in the SM. In general, such models
have a very large number of parameters, so that there is little
predictivity.

\end{enumerate}

Now, suppose that we find evidence for physics beyond the SM through the
measurements of CP asymmetries. How can we distinguish among the various
possibilities for new physics? Some progress can be made via a simple
observation. Any new physics which affects \bbbar\ mixing, which is a
flavour-changing process, will also affect rare flavour-changing
``penguin'' decays such as $b\to s X$ or $b\to d X$. For some models, or
regions of new-physics parameter space, the effects can be quite large. In
this case, the measurements of the branching ratios for penguin decays can
so constrain the parameters of the new physics as to render its effects in
\bbbar\ mixing, and hence the CP asymmetries, unimportant. It is an
experimental question whether or not measurements of the rates for such
penguin decays can be made before the CP asymmetries are measured.
Regardless, it is clear that measurements of CP asymmetries and penguin
decays will give complementary information. 

As an example, consider a model with $Z$-mediated FCNC's \cite{B-ZFCNC}.
The flavour-changing $Zb{\bar d}$ and $Zb{\bar s}$ couplings, which can
affect both \bbbar\ mixing and $B$ decays, are parametrized by $U_{db}$ and
$U_{sb}$, respectively. The present experimental bound on $BR(B \to
\mu^+\mu^- X)$ constrains these couplings to be
\beq
|U_{qb}| < 1.7 \times 10^{-3} ~.
\eeq
For maximal values of these parameters, \bdbdbar\ mixing may be dominated
by $Z$-mediated FCNC's; its contribution to \bsbsbar\ mixing can $\lsim
15$\%. In both cases, the CP asymmetries can be affected.

However, $Z$-mediated FCNC's also contribute to penguin decays. For
example, there is a tree-level contribution to the decay $b \to q l^+l^-$.
In the SM \cite{Ali},
\begin{eqnarray}
BR(B \to X_s \, \mu^+ \mu^-) \!\! &=& \!\! (5.7 \pm 1.3) \times 10^{-6}~, \nonumber
\\
BR(B \to X_d \, \mu^+ \mu^-) \!\! &=& \!\! (3.3 \pm 2.8) \times 10^{-7}~.
\end{eqnarray}
For maximal values of the $U_{qb}$ couplings, we find
\beq
BR(B \to X \, \mu^+ \mu^-) = 5 \times 10^{-5}~,
\eeq
which is 1-2 orders of magnitude above the SM prediction. (Of course, this
is a bit of a cheat, since $BR(B \to \mu^+\mu^- X)$ was used to constrain
the $U_{qb}$.)

Consider the decay $B_q^0 \to l^+ l^-$. For values of the decay constants
$f_{\sss B_s} = 232$ MeV and $f_{\sss B_d} = 200$ MeV, The SM predicts
\begin{eqnarray}
BR(B^0_s \to \mu^+ \mu^-) \!\! &=& \!\! (3.5 \pm 1.0) \times 10^{-9} ~, \nonumber \\
BR(B^0_d \to \mu^+ \mu^-) \!\! &=& \!\! (1.5 \pm 1.4) \times 10^{-10} ~. 
\end{eqnarray}
On the other hand, with $Z$-mediated FCNC's one finds
\begin{eqnarray}
BR(B^0_s \to \mu^+ \mu^-)|_{\sss Z-FCNC} \!\! &<& \!\! 5.8 \times 10^{-8} 
~, \nonumber \\
BR(B^0_d \to \mu^+ \mu^-)|_{\sss Z-FCNC} \!\! &<& \!\! 4.2 \times 10^{-8} 
~. 
\end{eqnarray}
Thus, for maximal values of the $U_{qb}$, the predicted rates for $B_s^0
\to l^+ l^-$ and $B_d^0 \to l^+ l^-$  are respectively about 20 and 300-400
times larger than those expected in the SM.

As a further example, consider electroweak penguin decays (EWP's), which
are mainly mediated by $Z$ exchange, rather than gluon exchange. An example
of such a decay, which involves the transition $b\to s$, is $\bs \to \phi
\pi^0$. Here the virtual $Z$ essentially turns into the $\pi^0$ -- because
of isospin, a gluon could not do this. $Z$-mediated FCNC's will of course
contribute to such decays. For this particular decay, we find
\beq
\left\vert {A_{\sss Z-FCNC} \over A_{\sss SM}} \right\vert 
< 5.5 ~.
\eeq
An example of an electroweak penguin decay involving a $b\to d$ transition
is $B^+ \to \phi \pi^+$. Here,
\beq
\left\vert {A_{\sss Z-FCNC} \over A_{\sss SM}} \right\vert
< 22.9 ~. 
\eeq
Obviously, the effects of $Z$-mediated FCNC's on such decays are enormous.
The branching ratios for pure electroweak penguin decays can be increased
by as much as a factor of $\sim 25$ ($b\to s$) or $\sim 500$ ($b\to d$)!
These are clearly ``smoking gun'' signals of new physics.

Another example of new physics which can significantly affect both \bbbar\
mixing and penguin decays is a fourth generation. Since the CKM matrix in
this case is $4 \times 4$, the parameter space is quite complicated.
However, suppose that $V_{td} \sim 0$, $V_{t'd} = 0.005$, $V_{tb} = V_{t'b}
\simeq 1/\sqrt{2}$, and $m_{t'} = 480$ GeV. Then \bdbdbar\ mixing is
dominated by the box diagram with internal $t'$ quarks. The phase of the
mixing is then $arg(V_{t'd}V_{t'b}^*)^2$, which may be quite different from
the SM. This will lead to CP asymmetries which may differ substantially
from the SM.

However, for this same choice of parameters, all penguin decays involving
the $b$-$d$ FCNC will also be dominated by the fourth generation. Comparing
the predictions for $b \to d$ penguin decays in this model with those of
the SM, we find
\begin{eqnarray}
BR( b \to d\gamma )|_{4-gen} \!\! &\simeq& \!\! 
{1\over 4} BR( b \to d\gamma )|_{\sss SM} , \\
BR( b \to d q{\bar q} )|_{4-gen} \!\! &\simeq& \!\! 
{1\over 5} BR( b \to d q{\bar q} )|_{\sss SM} , \nonumber \\
BR( \bd \to l^+ l^- )|_{4-gen} \!\! &=& \!\! 
8 \, BR( \bd \to l^+ l^- )|_{\sss SM} , \nonumber \\
BR( b \to d )_{\sss EWP} |_{4-gen} \!\! &=& \!\! 
8 \, BR( b \to d)_{\sss EWP} |_{\sss SM} . \nonumber
\end{eqnarray}
There are errors on the SM predictions, so the first two are only marginal
signals of new physics. However, the last two would be quite convincing
signals of physics beyond the SM.

Note that not all models of new physics which can affect \bbbar\ mixing,
and hence the CP asymmetries, have clear signals in penguin decays.
However, some of them do, so that measurements of CP asymmetries {\it and}
rare penguin decays will give complementary information. Both will be
necessary if we hope to identify the new physics.

\section{Conclusions}

To recap: in order to test the SM explanation of CP violation, it will be
necessary to observe it outside of the kaon system. There are numerous
possibilities for this. In this talk I have concentrated on three of them,
involving hyperons, neutral $D$ mesons, and neutral $B$ mesons. In all
cases, should CP violation be observed, we will want to know the answers to
three questions. Specifically, (i) can it be explained by the SM, (ii) if
not, what types of new physics can be responsible, and (iii) can we
distinguish among different models of new physics?

\begin{itemize}

\item CP Violation in Hyperon Decays: Such CP violation requires the
interference of tree and penguin diagrams, just like $\epsilon'/\epsilon$.
There are numerous processes and several CP-violating observables. Within
the SM, the asymmetries are small, of order $10^{-5}$. However, the
calculations have large theoretical uncertainties. In certain models of
physics beyond the SM, there may be enhancements in the CP asymmetries, but
these predictions also have large errors. In short, while it would be nice
to observe CP violation in hyperon decays, it will be very difficult to
determine if it is consistent with the SM, or if new physics is necessary.

\item Indirect CP Violation in $D^0$ Decays: Any such CP-violating
asymmetries require there to be \ddbar\ mixing. In the SM, this mixing is
negligible, so that there is no indirect CP violation. Going beyond the SM,
there are many models which can accomodate a mixing as large as the current
experimental limit. Note that the observation of such mixing would already
be a clear signal of new physics. If the mixing is sizeable, it may be
possible to also measure CP-violating asymmetries. The most promising
processes involve singly-Cabibbo-suppressed $D$ decays.

\item Indirect CP Violation in $B^0$ Decays: The SM predicts large
asymmetries in the neutral $B$ system. It is possible to extract CKM phase
information with no hadronic uncertainty. By measuring the angles and sides
of the unitarity triangle, it is possible to test the SM explanation of CP
violation. There are several ways in which new physics can manifest itself:
\begin{enumerate}

\item The relation $\alpha+\beta+\gamma=\pi$ is violated.

\item Although $\alpha+\beta+\gamma=\pi$, one finds values for the CP
phases which are outside of the SM predictions.

\item The CP angles measured are consistent with the SM predictions, and
add up to $180^\circ$, but are inconsistent with the measurements of the
{\it sides} of the unitarity triangle.

\end{enumerate}
In any of these cases, there are several models of physics beyond the SM
which could be involved. It may be possible to distinguish among the
different candidate models by looking at rare penguin decays. The
measurements of the CP asymmetries and such rare decays will give
complementary information.

\end{itemize}

\noindent
{\bf Acknowledgements}:

I am grateful to C. Kalman for the invitation to talk at this enjoyable
conference. I would also like to thank M. Gronau for his collaboration on a
number of the topics discussed here. This research was financially
supported by NSERC of Canada and FCAR du Qu\'ebec.



\end{document}